\documentstyle[aps,prd,epsfig]{revtex}
\newcommand{\bee}{\begin{equation}}
\newcommand{\ee}{\end{equation}}
\newcommand{\beea}{\begin{eqnarray}}
\newcommand{\eea}{\end{eqnarray}}

\preprint{COLO-HEP-467
}
\begin{document}

\title{Short Distance Current Correlators:
 Comparing Lattice Simulations to the Instanton Liquid}
\author{Thomas DeGrand}
\address{
Department of Physics,
University of Colorado, 
        Boulder, CO 80309 USA
}
\date{\today}
\maketitle
\begin{abstract}
Point to point
correlators of currents are computed in quenched QCD using a chiral lattice
fermion action, the overlap action. 
I compare correlators made of exact quark propagators with correlators 
restricted to low (less than 500 MeV) eigenvalue eigenmodes of the Dirac
operator. In many cases they show  qualitative resemblence
(typically at small values of the quark mass and distances larger
than 0.4 fm)
and they differ qualitatively at larger quark masses or at very short distance.
Lattice results are in qualitative agreement (and in the difference
of vector and axial vector channels, quantitative agreement)
with the expectations of
instanton liquid models. The scalar channel shows the effects of a
quenched finite volume zero mode artifact, a negative correlator.
\end{abstract}
%
%

\section{Introduction}

One of the earliest attempts to study nonperturbative behavior in QCD
was through an analysis of correlators of hadronic currents.
The QCD sum rule approach treats short distance physics in the
context of the operator product expansion (OPE)
and parameterizes long distance physics in terms of
 vacuum condensates\cite{ref:SVZ,ref:SURVEYSR}.
The approach was quite successful in the vector and axial vector channels,
but nonperturbative effects are very large in the pseudoscalar
 and scalar channels.
The realization that instanton effects appear in precisely those
 channels in which nonperturbative effects are large led to the
development of instanton liquid models\cite{ref:SS,ref:Diak}.
It would be an interesting exercise to ``validate'' the parameterization
and results of these models directly from the QCD Lagrangian,
by using lattice methods. This paper is a
 step in that direction.

We are concerned with flavor non-singlet current correlators
separated by a (four dimensional Euclidean) distance $x$,
\bee
\Pi_i(x) = {\rm Tr}\langle J_i^a(x)J_i^a(0)\rangle
\ee
where the current will be proportional to
\bee
J_i^a(x) = \bar \psi(x) {\tau^a } \Gamma(i) \psi(x)
\ee
with $\tau^a$ an isospin label and $\Gamma$ a product of Dirac matrices.
 In practice (and to be consistent with the standard approach to
these correlation functions\cite{ref:s93})
 I present results for the correlators scaled
with respect to the massless
 free field current-current correlator $\Pi_i^0(x)$
 (the precise kind of free fermions will be described below)
\bee
R_i(x) = \Pi_i(x)/\Pi_i^0(x).
\label{eq:ratio}
\ee
The label $i$ will include the pseudoscalar (PS), scalar (S), vector (V)
and axial vector (A) currents.

The Euclidean space correlators whose Fourier transforms into coordinate space
I am  measuring obey a dispersion relation which relates them to
their absorptive part in Minkowski space,
the spectral density $\rho(s)$, 
\bee
\Pi(q^2) = {1\over \pi}\int ds {{\rho(s)}\over{s+q^2}}.
\ee
$\rho(s)$ is 
in turn is proportional
to the total cross section for scattering in the appropriate channel,
and should then be positive. The point-to-point correlators measured here are
then related to the spectral density through
\bee
\Pi(x) = \int  d^4 q \exp(iq\cdot x) {1\over \pi}
 \int ds {{\rho(s)}\over{s+q^2}}
\ee
or, more directly,
\bee
\Pi(x) ={1\over \pi} \int ds \rho(s) D(\sqrt{s},|x|)
\ee
where
\bee
D(\sqrt{s},|x|)= {1\over{4\pi^2 }} \int_0^\infty
{{p^2 d p} \over {\sqrt{p^2+s}}} \exp(-|x|\sqrt{p^2+s})
\label{DX}
\ee
is the Euclidean propagator for a free particle of squared mass $s$.
If we were measuring a conventional (for the lattice) $\vec q=0$ correlator,
Eq. \ref{DX} would be replaced by
\bee
D_{\vec q=0}(m,|x|) = { {\exp(-m|x|)}\over{2m} }.
\ee

In principle, lattice calculations could check
results of instanton liquid models\cite{ref:s93}. In
practice, however, lattice simulations have been contaminated by the
use of fermion actions which are not chiral. Since the continuum
analyses depend crucially on chiral symmetry arguments, one might
wish to be cautious about drawing conclusions from simulations with
 nonchiral lattice fermions.  The recent discovery
of lattice actions which support an exact chiral symmetry\cite{ref:GW}
(notably the overlap action\cite{ref:neuberfer})
 allows one to revisit these questions
 in a theoretically clean context.
That is the subject of this paper: a study of current correlators from an
overlap fermion action.

Because I am working with an action with exact chiral symmetry,
 I have to be careful with my
definitions and make them consistent with current commutation
relations. If I assume that I am
 modeling a theory with a flavor $SU(2)_V \times SU(2)_A \simeq O(4)$
symmetry, I can write down two real $O(4)$ vectors of currents
\bee
\vec \phi_1 = (\pi^a,f_0)
\ee
and
\bee
\vec \phi_2 = (a_0^a,\eta)
\ee
where
 $\pi^a= i \bar\psi \gamma_5 (\tau^a) \psi \equiv J_{PS}$
 is the pseudoscalar (PS) current,
 $a_0^a= -\bar\psi (\tau^a/2) \psi\equiv J_S$
 is the scalar (S) isotriplet current,
 $f_0 = \bar\psi  \psi$ is the scalar isoscalar  current, and
 $\eta = i \bar\psi \gamma_5 \psi$ is the pseudoscalar isoscalar current.
(Correlators of the last two currents involve disconnected diagrams and I will
not consider them further.)

In the cases of the  V (vector) ($\Gamma = \gamma_\mu$),
and
 A (axial vector) ($\Gamma = \gamma_\mu \gamma_5$) currents,
I sum the correlator over the four values
of $\mu$.
The most interesting way to present results for correlators in the
vector and axial vector channels is to look at the sum or difference of
vector and axial vector correlators
\bee
R_{V \pm A}(x)= {{\Pi_V(x) \pm \Pi_A(x) }\over{2 \Pi_V^0(x)}}.
\ee
In the sum rule/OPE approach
$R_{V + A}$ is dominated by perturbative physics and is expected
to take a value very close to unity,
while $R_{V - A}$ is zero at small $x$ and receives only nonperturbative
contributions which are relevant to chiral symmetry breaking. 
The instanton liquid model produces a large part of the correlator
$R_{V - A}$, 
as emphasized recently by Schafer and Shuryak \cite{ref:ssa1}.

In a recent publication \cite{ref:dh00} we have shown that the
low lying eigenmodes of a chiral lattice fermion Dirac operator,
an overlap fermion operator (described in Ref. \cite{ref:TOM_OVER}),
has a local chiral density
 $\psi(x)^\dagger \gamma_5 \psi(x)$ which shows a peaked structure.
The  positions and signs of the peaks
 are strongly correlated with the locations of
topological objects, which would be identified as
 instantons and anti-instantons detected using a pure gauge operator.
Zero modes correlate with only one sign of topological objects, while nonzero
eigenmodes of the Dirac operator interpolate between both signs of topological
object.  This correlation dies away slowly as the eigenvalue of the mode
rises.  Spatially averaged correlation functions of hadrons made of light
quarks are saturated by propagators of quarks restricted to a few low
eigenmodes.
The whole picture is very reminiscent of an instanton liquid model.

Ref. \cite{ref:dh00} did not consider the possibility that the density
of the fermionic modes might be large in places where the local chiral
density was small. Subsequently, this point was raised by the
authors of Ref. \cite{ref:horv01}. Ref. \cite{ref:dh01} showed that
this possibility did in fact not occur for the overlap action used in Ref.
\cite{ref:dh00}.
 Similar results have been presented using an alternative chiral
lattice action (the Wilson overlap action) \cite{ref:he01}
two lattice actions with improved but inexact chiral
 symmetry (domain wall fermions \cite{ref:bl01} and an approximate
Ginsparg-Wilson action \cite{ref:regen})
and with an improved operator in a lattice action with inexact chiral symmetry
(the clover action) \cite{ref:hi01}.
 None of these last three works compared the
distribution of chirality with topological charge density measured with a
gauge observable.



It is an uncontrolled approximation to replace a quark propagator by
the quark propagator restricted to a sum over a small number of
small eigenvalue eigenmodes
of the Dirac operator. However, if an observable computed with truncated
propagators resembles the same observable computed with exact propagators,
and if we  further have seen that the low modes couple to particular structures
in the QCD vacuum, then this resemblence
is a strong (qualitative) signal that the particular vacuum structure
is connected with that observable.

For example, consider the difference of vector and axial vector currents.
$R_{V-A}(x)$, either extracted from $\tau-$ decay or measured in Monte Carlo
simulation, is zero at small $x$ and increases with $x$. The same
behavior is seen in a calculation based on propagators truncated to
a small number of eigenmodes: in fact, at small quark mass (pseudoscalar to
vector mass ratio less than 0.5) the difference between the exact and
truncated $R_{V-A}(x)$ is small.
Since in \cite{ref:dh00} the locations of peaks in the low eigenvalue modes
were correlated with the locations of peaks in the topological charge density,
one can infer that instantons are connected with the rise in $R_{V-A}(x)$.

If the current correlator based on truncated propagators does not
reproduce the full calculation, the truncated propagator must be missing some
 important physics.
For example, in most channels, $R(x)$ approaches unity as $x$ falls to zero.
 Approximating quark propagators by a few low
eigenvalue modes
does not reproduce this behavior: in that case $R(x)$ typically vanishes
at small $x$.
Presumably $R(x)\rightarrow 1$ as $x \rightarrow 0$
 is asymptotic freedom at work: the correlator is reducing
 to its free-field value. The low eigenmodes are extended in space and
decouple from short distance physics.

Previous studies\cite{ref:chu93,ref:hands95}
 of point-to-point correlators have mostly been concerned
with the long-distance behavior of these observables, including
the extraction of particle masses and couplings. I will not attempt
to do that here, because my simulation volume is very small.
Some studies\cite{ref:IN} have also compared the correlators in a ``full'' QCD
simulation with propagators truncated to a set of
 low-eigenvalue eigenmodes. They see results similar to the ones I report,
 the main difference
being that they need many more eigenmodes to saturate the correlator.
I believe that this difference is due to the poor chiral behavior of
 the action, the usual thin link Wilson action, used by these authors.

Along the way, I observe an interesting artifact of the quenched approximation
in the scalar channel--the correlation function becomes negative.
This behavior is incompatible with a normal spectral representation.
In my data it seems to be associated with a finite-volume quenched 
approximation artifact arising from the exact zero modes of the Dirac operator.

\section{The Lattice Calculation}
The overlap action used in these studies\cite{ref:TOM_OVER}
 is built from an action with nearest and
next-nearest neighbor couplings,
and  APE-blocked links\cite{ref:APEblock}.
Eigenmodes of the massless
 overlap Dirac operator $D(0)$ are constructed from eigenmodes
of the Hermitian Dirac operator $H(0)=\gamma_5 D(0)$,
 using an adaptation of
a Conjugate Gradient algorithm of Bunk et. al.
and  Kalkreuter and Simma\cite{ref:eigen}.
These eigenmodes are used to precondition
the calculation of the quark propagator (by use of a conjugate gradient
algorithm) and are additionally used to construct quark propagators
truncated to some number of low lying eigenmodes.

Later, we
 will need a few simple facts about the eigenmodes of the overlap Dirac
operator, so let us recall them now.
 The eigenmodes of any massless
overlap operator are located on a circle in the complex plane of radius $x_0$
with a center at the point $(x_0,0)$. The corresponding eigenfunctions are
either chiral (for the eigenmodes with real eigenvalues
located  at $\lambda=0$ or $\lambda=2x_0$)
or nonchiral  and paired; the two eigenvalues of the nonchiral modes
are complex conjugates.
 The massive overlap Dirac operator is
conventionally defined to be 
\bee
D(m) = ({1-{m \over{2x_0}}})D(0) + m
\ee
and it is also conventional to define the propagator so that the chiral
modes at $\lambda=2x_0$ are projected out,
\bee
\hat D^{-1}(m) = {1 \over {1-m/(2x_0)}}(D^{-1}(m) - {1\over {2x_0}}) .
\ee
Then the contribution to the propagator of a single (positive chirality)
zero mode in the basis
where $\gamma_5= {\rm diag}(1,-1)$ is
\bee
\hat D(m)^{-1}= {1\over m}\pmatrix{1&0\cr 0 & 0 \cr}.
\ee
The $j$th pair of nonchiral modes contributes a term
\bee
\hat D(m)_j^{-1}= \pmatrix{ \alpha_j& -\beta_j\cr
 \beta_j & \alpha_j \cr},
\ee
where, defining $\mu=m/(2x_0)$, $\epsilon_j= \lambda_j/(2x_0)$, the entries are
\bee
\alpha_j={1\over{2x_0}}
{{\mu(1-\epsilon_j^2) }\over{\epsilon_j^2 + \mu^2(1-\epsilon_j^2)}}
\ee
\bee
\beta_j={1\over{2x_0}}
{{\epsilon_j\sqrt{1-\epsilon_j^2})  }\over
{\epsilon_j^2 + \mu^2(1-\epsilon_j^2)}}
\ee
and $D(0)^2\phi_j= \lambda_j^2\phi_j$; the eigenmodes of $D(0)$
have eigenvalues $2x_0(\epsilon_j^2 \pm i\epsilon_j\sqrt{1-\epsilon_j^2})$.
For a summary of these useful formulas, see Ref. \cite{ref:FSU98}
(for the special case $x_0=1/2$).

The data set used in this analysis are generated in quenched approximation
 using the Wilson gauge action at a coupling
 $\beta=5.9$. The nominal lattice spacing is $a=0.11$
 (inferred from the Sommer parameter
using the interpolation formula of Ref. \cite{ref:precis}) or 0.13 fm
from the rho mass. This will be discussed in more detail below.
 It consists of 20 $12^4$
configurations. The fermions have periodic boundary conditions in the
 spatial directions and anti-periodic temporal boundary conditions.
I calculated the ten  smallest eigenvalue modes of $H^2(0)$ in the chiral
sector of the  minimum eigenvalue,
 and reconstructed the degenerate eigenstate
of opposite chirality of $H^2(0)$ when one was present.  These modes are
then recoupled into eigenmodes of $D(0)$.
Their eigenvalues have imaginary parts ranging up to $0.3/a$-$0.35/a$,
or about 500 MeV \cite{ref:dh00}.

The lattice analysis has one strong point and one weak point.
The strong point is the use of an action with exact chiral symmetry.
This means that there is no additive renormalization of the quark mass.
It also means that there are no exceptional configurations, so the simulations
can be performed at small quark mass. Instead of exceptional configurations,
there are contributions to the quark propagator from zero modes of
the Dirac operator. Because of chiral symmetry, these zero modes 
contribute in a ``dangerous'' way, as finite volume effects
only in the PS and S channels.
The particular choice of action, with a fat link,
 also has rather mild multiplicative renormalization of current
matrix elements. The perturbation theory has not been done for this action, but
similar calculations done for the fat link clover action have vector
and axial vector lattice-to-continuum renormalization factors quite close
to unity\cite{ref:BD}.

The weak point of the calculation is its use of a small volume. The
lattice has a size of about 1.5 fm, assuming a lattice spacing of
0.13 fm. 

One annoyance encountered in this study is the different finite volume
effects suffered by the free massless current correlator and the
correlator in the nontrivial gauge background. At large distance the
free correlator can receive contributions in which the quark and antiquark
wind in opposite directions around the lattice, so that the correlator has
 the topology of a single line encircling the simulation volume. This
contribution is not present in the nontrivial background, since it corresponds
to a single Polyakov loop winding around the lattice.
Confinement forces the expectation value of this operator to be zero.
 The way I got around
this problem is the same as was used in Ref. \cite{ref:chu93}:

Think of the finite volume lattice as a piece of an infinite lattice.
A propagator from a point source in the finite lattice corresponds, on the
infinite lattice, to
a sum of propagators from sources which include the original source, plus
a set of image points on all the ``copies'' of the finite lattice which tile
the infinite lattice.  We only want to include contributions to correlators
from infinite-volume propagators
where the source points for the quark propagator and antiquark
propagator coincide, and so do the sink points. These
 quark-antiquark  correlators
look like  closed fermion loops. Contributions in the finite volume in 
which the quark and antiquark propagate in opposite directions around the 
finite volume to get to the same sink
point are topologically equivalent to lines in the infinite volume,
where the quark source and antiquark source sit on different image points
and the two propagators terminate on the same sink point.

To compute the necessary free field current-current correlator, I
construct an approximation to the infinite volume free fermion propagator by
computing the free (overlap) fermion propagator
 on a large lattice (I used a $24^4$ volume) and
approximate the quark-antiquark correlators in the small volume
 as a sum of the ``direct''
term and propagators from sources located on (the same)
  nearest image point.

The hadron correlators also suffer finite volume effects. In particular,
the lattice volume cuts off the long distance part of the hadron correlator,
which in most cases contains the contribution to the correlator of the lightest
state in the channel. Thus, I have not tried to do any fits to the mass
of this state, in contrast to Refs. \cite{ref:chu93,ref:hands95}.
Note however, that for short distances (less than 0.5 fm or so),
the effects of finite volume are not too important. Also, comparisons
of a full correlator to a correlator computed using truncated propagators
are not affected by the free field correlators, so the observation that
low modes saturate a correlator (or not) are not so compromised.

At very light quark mass, chiral symmetry breaking begins to be modified by
the small volume. The relevant parameter is the Leutwyler-Smilga
\cite{ref:Leutwyler}
parameter, $z= m_q \Sigma V$ for quark mass $m_q$, infinite volume
 condensate $\Sigma=\langle \bar q q \rangle$, and simulation volume $V$.
A large volume corresponds to $z \gg 1$.
Using the value of $\Sigma$ computed for this action at this coupling
in Ref. \cite{ref:TOM_OVER}, $z \simeq0.8 (a m_q/0.01))$ (to show $z$ in
units of the smallest quark mass used). At this quark mass the scaled
lattice size $L$ is $m_\pi L \simeq 2.4$.

\section{Results}

\subsection{The pseudoscalar and scalar channels}

\begin{figure}[thb]
\begin{center}
\epsfxsize=0.8 \hsize
\epsffile{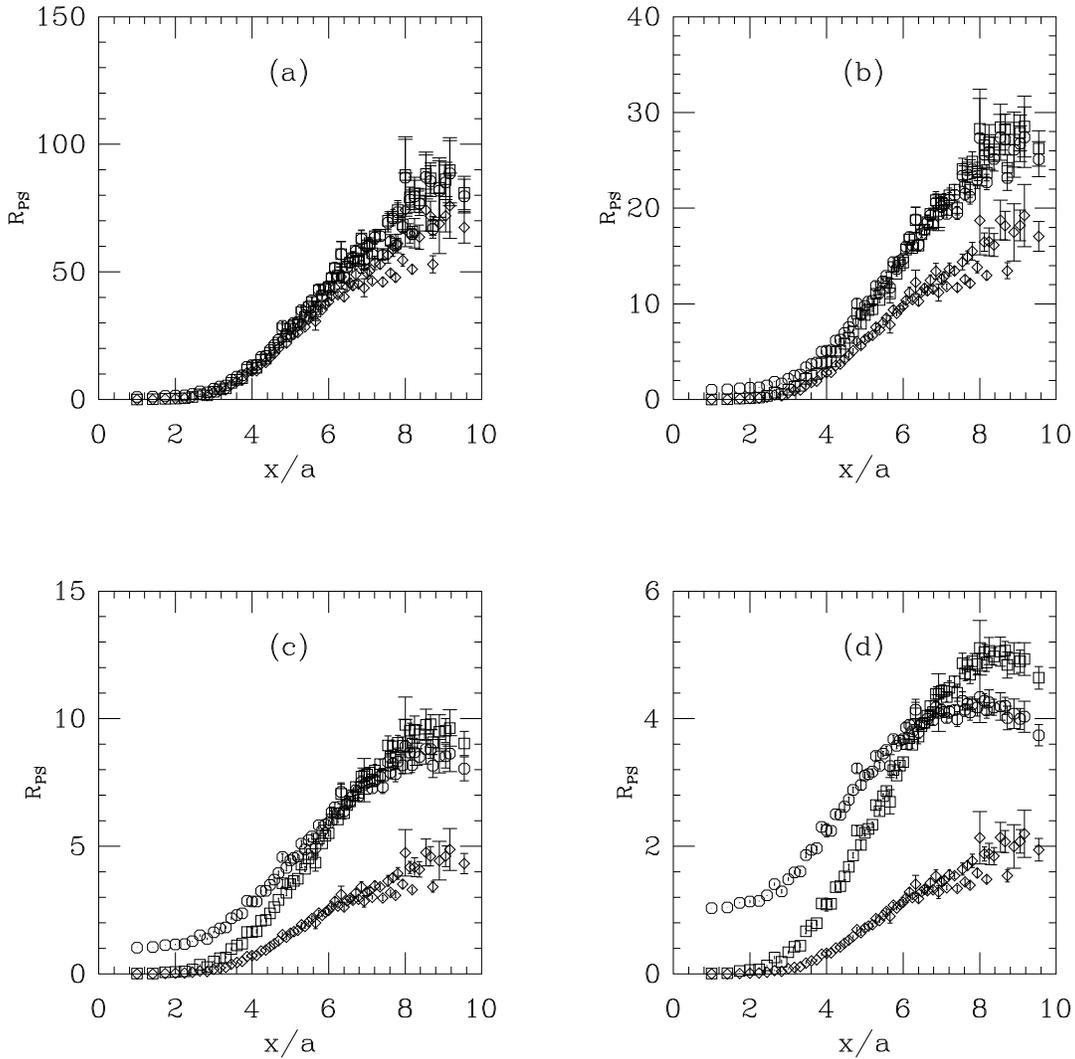}
\end{center}
\caption{
Saturation of the point-to-point pseudoscalar correlator by 
 low-lying eigenmodes of $H(0)^2$.
 (a) $am_q=0.01$ ($m_{PS}/m_V\simeq 0.34$);
 (b) $am_q=0.02$ ($m_{PS}/m_V\simeq 0.50$);
 (c) $am_q=0.04$ ($m_{PS}/m_V\simeq 0.61$);
 (d) $am_q=0.06$ ($m_{PS}/m_V\simeq 0.64$).
 Octagons  show the full hadron correlator.
Squares show the contribution from the lowest 10 modes.
 Diamonds show the contribution from the zero modes,
 which just scales as $1/(am_q)^2$).
}
\label{fig:pseudo}
\end{figure}

 In Fig. \ref{fig:pseudo} is shown the pseudoscalar
correlator for several choices of light quark masses, with pseudoscalar to
vector meson mass ratios ($m_{PS}/m_V$) inferred from Ref. \cite{ref:dh00}:
 (a) $am_q=0.01$ ($m_{PS}/m_V\simeq 0.34$);
 (b) $am_q=0.02$ ($m_{PS}/m_V\simeq 0.50$);
 (c) $am_q=0.04$ ($m_{PS}/m_V\simeq 0.61$);
 (d) $am_q=0.06$ ($m_{PS}/m_V\simeq 0.64$).
The separation $x$ is measured in units of the lattice spacing in the
simulation.
Also shown is the contribution to the correlator from the lowest ten modes of
the Dirac operator, as well as the contribution to the correlator
given by the zero modes alone.
Quark interactions in this channel are strongly attractive.
The enormous attraction in the pseudoscalar channel is well known in
continuum (compare Ref. \cite{ref:Novikov})
and lattice (compare Ref. \cite{ref:chu93})
analyses. In instanton liquid phenomenology it is ascribed to the effect of
the `t Hooft interaction, which is strong and attractive in the
pseudoscalar channel.
At low quark masses ($m_{PV}/m_V \le 0.5$ or so) the low mode truncation
saturates the correlator at larger $x$.
 As these are
the modes which couple to instantons, there seems to be a connection
between instantons and the strong attraction in the pseudoscalar channel
at $x/a = 5-9$.

The fact that the contribution of the low modes begins at $R(x=0)=0$ and not 1
is just a sign that at tiny $x$ the channel is dominated by free-field modes
(which presumably do not couple to topology), but by a distance
of 2-4 lattice spacings (0.2-0.4 fm) the light quark mass
correlator is completely dominated
by the low modes.
As the quark mass falls to zero, the pure zero modes make an ever larger
contribution to the point-to-point correlators, as well: their
contribution scales a $1/m_q^2$ times a quark-mass-independent function
of $x$ and $y$,
\bee
\Pi^{zero}(x,y) = {1 \over m_q^2}
 | \sum_{j \in zero}\phi_j(x)\phi_j(y)^\dagger|^2.
\ee

\begin{figure}[thb]
\begin{center}
\epsfxsize=0.8 \hsize
\epsffile{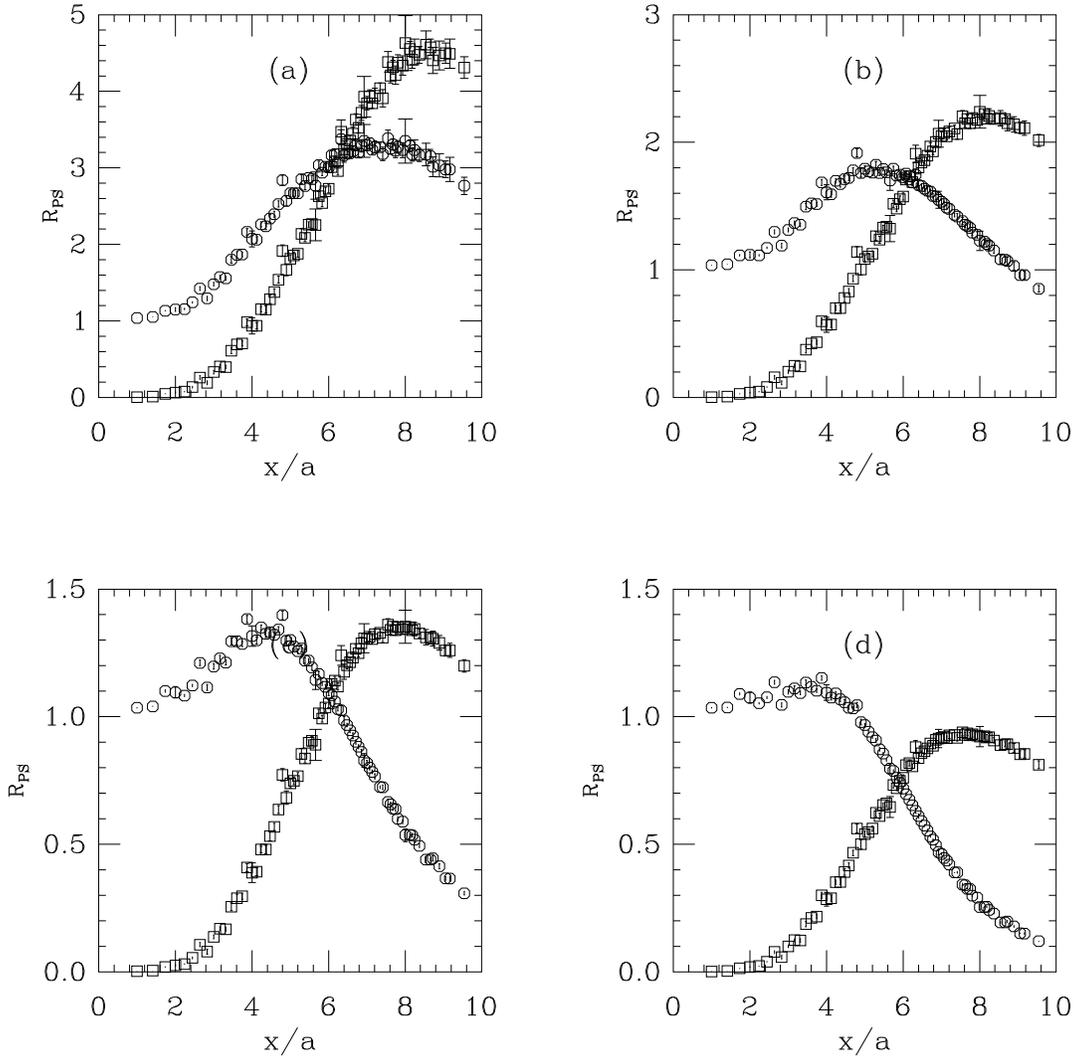}
\end{center}
\caption{
Non-saturation of the point-to-point pseudoscalar correlator for heavier
quark masses by  low-lying
eigenmodes of $H(0)^2$.
 (a) $am_q=0.10$ ($m_{PS}/m_V\simeq 0.75$);
 (b) $am_q=0.15$ ($m_{PS}/m_V\simeq 0.84$);
 (c) $am_q=0.20$ ($m_{PS}/m_V\simeq 0.87$);
 (d) $am_q=0.25$ ($m_{PS}/m_V\simeq 0.91$).
 Octagons  show the full hadron correlator.
Squares show the contribution from the lowest 10 modes.
}
\label{fig:pseudoh}
\end{figure}

As the quark mass rises, the approximation of $R_{PS}(x)$ 
by a small number (ten) of low-lying eigenmodes
becomes poorer and poorer.
This is shown in Fig. \ref{fig:pseudoh}. The correlators flatten and approach
unity over a wide range of  $x$.
That the low eigenvalue modes do not saturate the correlator
shows physics of these heavier pseudoscalars has
little to do with fermion eigenmodes which are strongly coupled to instantons.
This is not a surprise. It points up the dangerous possibility, however, that
lattice simulations involving pseudoscalars performed at a pseudoscalar
to vector ratio above 0.7 or so might give misleading results when extrapolated
to small quark mass.

\begin{figure}[thb]
\begin{center}
\epsfxsize=0.8 \hsize
\epsffile{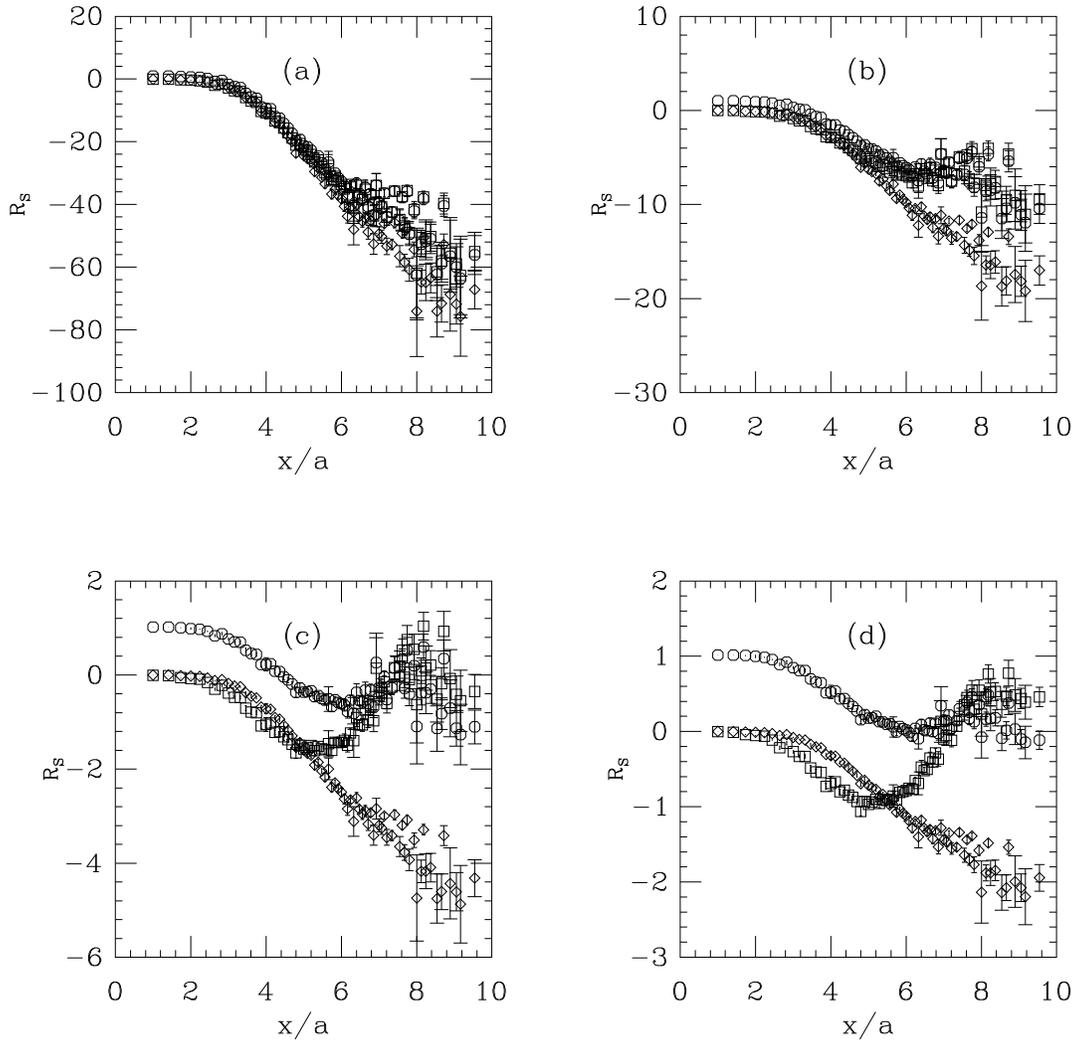}
\end{center}
\caption{
Saturation of the point-to-point scalar correlator by 
 low-lying eigenmodes of $H(0)^2$.
 (a) $am_q=0.01$ ($m_{PS}/m_V\simeq 0.34$);
 (b) $am_q=0.02$ ($m_{PS}/m_V\simeq 0.50$);
 (c) $am_q=0.04$ ($m_{PS}/m_V\simeq 0.61$);
 (d) $am_q=0.06$ ($m_{PS}/m_V\simeq 0.64$).
 Octagons  show the full hadron correlator.
Squares show the contribution from the lowest 10 modes.
 Diamonds show the contribution from the zero modes,
which again just scales as $1/(am_q)^2$.
}
\label{fig:scalar}
\end{figure}

The scalar correlator shows similar qualitative behavior to the
pseudoscalar correlator: strong deviation from unity away from  small $x$
and at small quark mass a large contribution from zero modes.
 See Fig. \ref{fig:scalar}.

Instanton liquid models predict a repulsive interaction in this channel.
We see that, indeed, the interaction between quarks is strongly repulsive --
so strongly repulsive, that at
small quark mass, the correlator becomes negative at larger $x$.

The bulk of this effect is due to the zero modes.
To make the argument, I begin by considering
the susceptibilities,
or spatial integrals of the correlators \cite{ref:FSU98}:
 In the pseudoscalar channel,
configuration by configuration,
 if the configuration has topological charge $Q$,
\bee
\chi_\pi =   {1 \over V} \sum_{x,y,a}\langle J_{PS}^a(x) J_{PS}^a(y)\rangle
           = {2 \over V} {\rm Tr} \ (\hat D^{-1} \gamma_5)^2 
\ee
\bee
 = 2 {{|Q|}\over{m^2 V}} + {4 \over V} \sum_j (\beta_j^2 + \alpha_j^2  )
\ee
\bee
=  {2 \over m} \langle \bar \psi \psi \rangle
\ee
The last line is the Gell-Mann, Oakes, Renner relation,
which is satisfied exactly by overlap fermions.  The isotriplet
scalar susceptibility is
\bee
\chi_{a_0} =  {1 \over V} \sum_{x,y,a}\langle J_{S}^a(x) J_{S}^a(y)\rangle 
= -{2 \over V} {\rm Tr} \ \hat D^{-2}
\ee
\bee
  = 
-2 {{|Q|}\over{m^2 V}} + {4 \over V} \sum_j ( \beta_j^2 - \alpha_j^2).
\ee
Note the negative contribution of the zero modes to the scalar channel.
In general, one would expect the sum over the nonzero modes to contribute
a result of order unity to the susceptibility. The contribution of the
 zero modes should scale away as $1/\sqrt{V}$, since the existence of
a  topological susceptibility means that $\langle Q^2 \rangle \simeq V$.
Thus the zero mode contribution to both the pseudoscalar and scalar channels
 is a finite volume artifact, but one
which dies away quite slowly.

So much for the integral of the correlator; what about its
value point by point? The contributions where  both of the quark
lines propagate through zero modes are equal in magnitude and opposite in sign
in the pseudoscalar and scalar channels. If the correlator is dominated
by modes whose eigenvalues are large compared to the quark mass, one
would expect the nonzero mode sum to contribute positively to the correlator.
This is certainly the case for the free theory, and so we naturally have
a situation where the free scalar correlator is positive and
the interacting correlator is dominated by its zero modes and is negative.

\begin{figure}[thb]
\begin{center}
\epsfxsize=0.8 \hsize
\epsffile{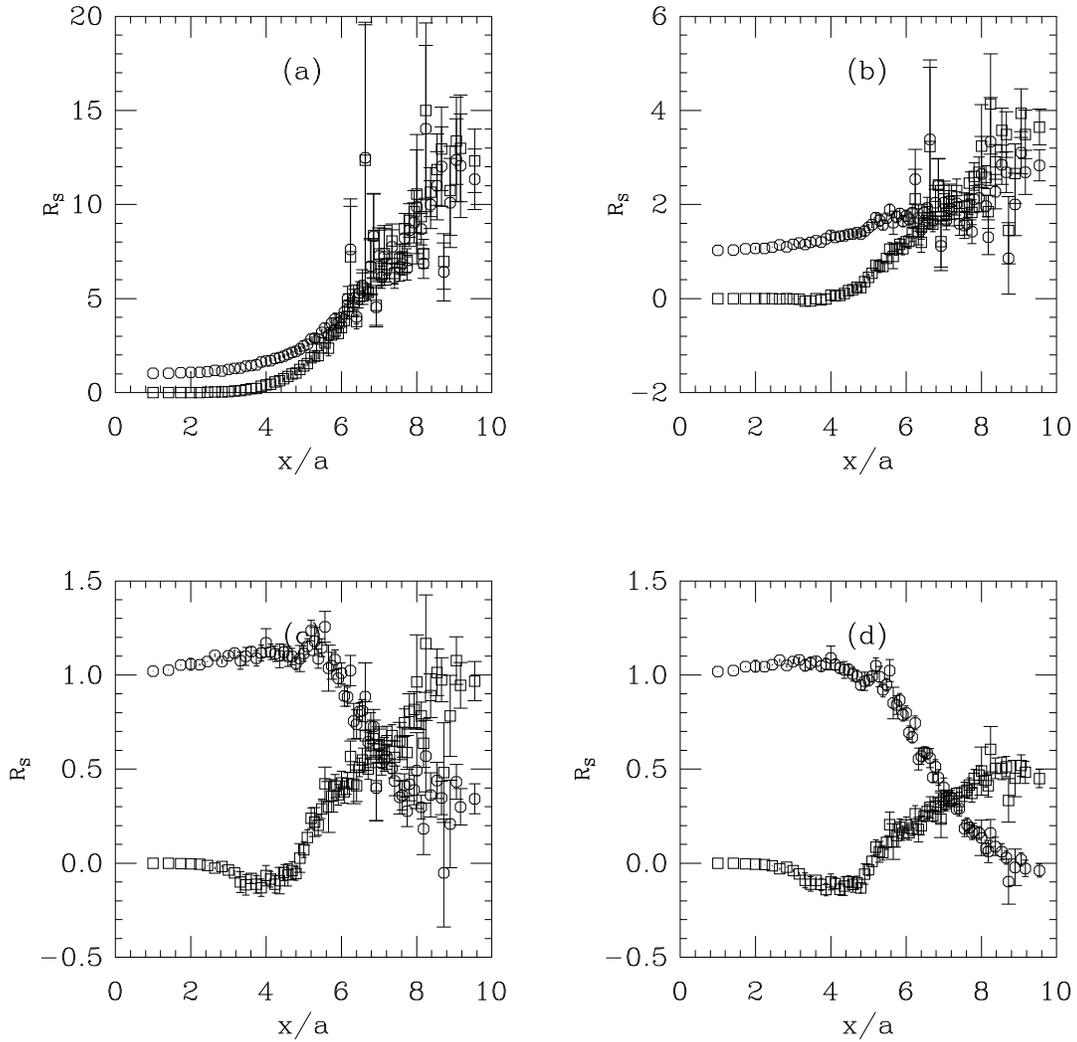}
\end{center}
\caption{
Comparison of the point-to-point scalar correlator to the scalar correlator
composed of low-eigenvalue quark modes in the $Q=0$ sector.
 (a) $am_q=0.01$ ($m_{PS}/m_V\simeq 0.34$);
 (b) $am_q=0.02$ ($m_{PS}/m_V\simeq 0.50$);
 (c) $am_q=0.04$ ($m_{PS}/m_V\simeq 0.61$);
 (d) $am_q=0.06$ ($m_{PS}/m_V\simeq 0.64$).
 Octagons  show the full hadron correlator.
Squares show the contribution from the lowest 10 modes.
}
\label{fig:nscalar}
\end{figure}

One can check that the zero mode is the source of the loss
 of spectral positivity by considering the scalar propagator in the
subset of our lattices which have zero topological charge (5 of the 20
lattices). This is shown in Fig. \ref{fig:nscalar}. Notice that
the scalar correlator in this sector is still saturated by low eigenmodes
 of the Dirac operator for $m_{PS}/m_V \le 0.5$ at a distance of 5 lattice
spacings ($x \simeq 0.55$ fm). In this sector the channel
 only becomes strongly repulsive at heavier quark masses, where the low
 modes do not contribute to the correlator.

The $Q=0$ sector of QCD is not QCD and so it remains an open question, 
what will happen at larger volumes.

Finally, one might worry that all of the attraction in the pseudoscalar channel
is due to the zero modes. The sum of pseudoscalar and scalar correlators,
\bee
R_{PS+S}(x)= {{\Pi_{PS}(x) + \Pi_S(x) }\over{2 \Pi_{PS}^0(x)}}.
\ee
has no zero mode contribution.

\begin{figure}[thb]
\begin{center}
\epsfxsize=0.8 \hsize
\epsffile{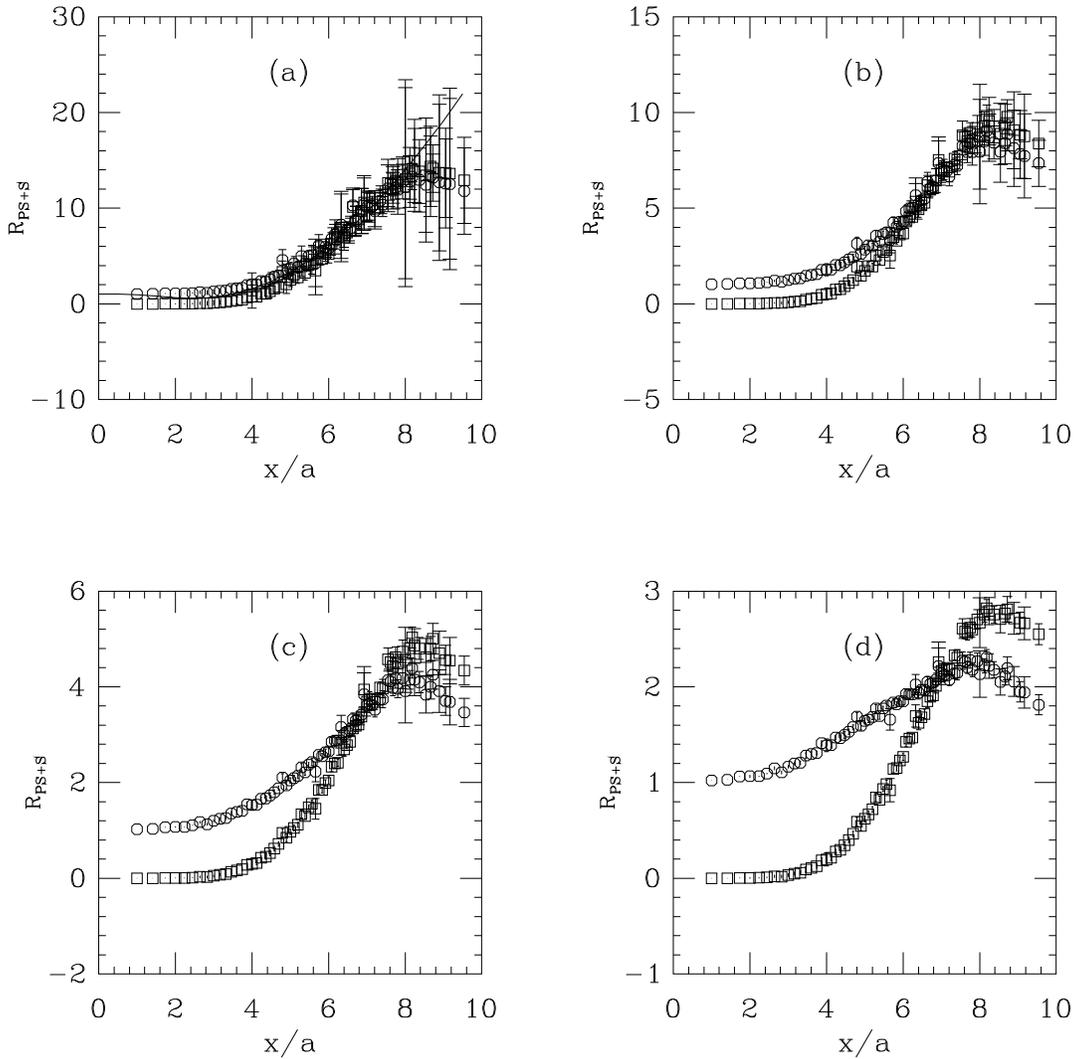}
\end{center}
\caption{
Saturation of the sum of  point-to-point pseudoscalar and
 scalar correlators sector by 
 low-lying eigenmodes of $H(0)^2$.
 (a) $am_q=0.01$ ($m_{PS}/m_V\simeq 0.34$);
 (b) $am_q=0.02$ ($m_{PS}/m_V\simeq 0.50$);
 (c) $am_q=0.04$ ($m_{PS}/m_V\simeq 0.61$);
 (d) $am_q=0.06$ ($m_{PS}/m_V\simeq 0.64$).
 Octagons  show the full hadron correlator.
Squares show the contribution from the lowest 10 modes.
The curve in panel (a) is from a simple phenomenological parameterization
of the spectral density.
}
\label{fig:psscalar}
\end{figure}

This quantity is shown in Fig. \ref{fig:psscalar}.
 The low modes do not contribute to the correlator at small $x$. At larger
$x$ they show a significant attractive interaction, saturating the
full-propagator correlator at larger $x$ for $m_{PS}/m_V \le 0.5$.
At heavier masses they produce too much attraction at larger $x$.

Other possibilities for a negative spectral weight could exist in the scalar
 channel.
Schafer and Shuryak \cite{ref:ss96} actually show that the scalar correlator
 in interacting instanton models becomes negative at $x \simeq 0.5$ fm.
They ascribe this behavior to the  strong repulsion in the scalar channel
expected in instanton models becoming a little too strong, and their
calculations of interacting instantons which include dynamical fermion
effects soften the interaction and restore positivity.

\begin{figure}[thb]
\begin{center}
\epsfxsize=0.4 \hsize
\epsffile{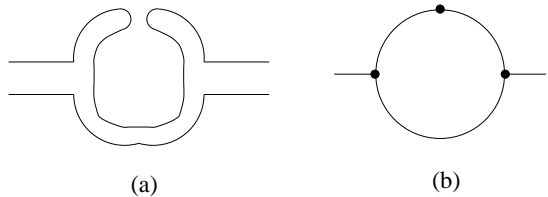}
\end{center}
\caption{
The quark line graph (a) and associated Feynman graph (b)
 which makes a ghost contribution to the scalar correlator.
}
\label{fig:hairpin}
\end{figure}

Recently, a discussion of the same effect has been given by
Thacker {\it et al}\cite{ref:thacker} in the context of quenched chiral
 perturbation theory\cite{ref:chiralpt}:
In full QCD the quark and antiquarks in the
eta-prime, the flavor singlet pseudoscalar meson, can annihilate and the
(multi) $q\bar q$ pair(s)  Fock state of the meson can mix with a quarkless
intermediate state. This mixing shifts the mass of the eta-prime.
In the quenched approximation the sum of mixing graphs truncates with
a single term  and the propagator
consists of a single pole and a double pole with a negative coefficient,
a ghost.
\bee
G(p) = {1  \over{ p^2+m^2}} - {1 \over {p^2+m^2}}{ \mu^2 \over 3 }
{1 \over {p^2+m^2}}
\label{ghost}
\ee
A negative spectral weight in the scalar channel can appear because
the scalar particle can couple to an intermediate state which is
a combination of a pion and a quenched eta-prime, whose propagator is
the second term (hairpin) part of Eq. \ref{ghost}.
This is a connected graph so in the quenched approximation it
can be constructed. It is shown in Fig. \ref{fig:hairpin}.

While we do not know the overall magnitude
of the graph, the particle in all the propagators is the flavor nonsinglet
pseudoscalar, whose mass is known from conventional spectroscopy calculations.
If I  assume that the scalar particle-pion-eta coupling  is just
a scalar constant $g$, then I can compute $\rho(s)$ from the graph of
Fig.\ref{fig:hairpin}(b). 
Because of the minus sign in Eq. \ref{ghost}
 the amplitude is negative
and it is also divergent at threshold,
\bee
 \rho(s) = -{{ \mu^2 g^2} \over{\sqrt{s(s-4m_\pi^2)}}}.
\label{imghost}
\ee
Then the shape of $R(x)$ can be computed exactly up to an overall
 normalization.
This contribution can have nothing to do with the zero modes because the
spatial correlations of the zero modes do not depend on the quark mass.
Attempts to fit the lattice data of Fig. \ref{fig:scalar} to a 
conventional \cite{ref:s93} 
combination
of continuum-like background $\rho(s)= (3s/(8\pi^2))\theta(s-s_0)$ 
plus a possible scalar resonance plus the ghost term (even varying the
strength of the ghost coupling with quark mass)  were unsuccessful.

Sch\"afer and Shuryak have presented predictions for point to point
correlators in the instanton liquid model. It would be an interesting
exercise to compare lattice results to theirs. Because of the presence
of the zero mode contribution in the pseudoscalar and scalar channels, it 
probably makes the most sense to consider the sum of pseudoscalar and
scalar channels, from which the zero modes decouple. Next, one must
choose a lattice spacing. Because the correlators rise up so strongly with
$x$, they are very sensitive to the choice of lattice spacing.
 I have
collected a number of observables for this simulation, which happen to have
more or less well-known continuum values, and which might be used to
infer a lattice spacing, in Table \ref{tab:latts}.
 The quantity $r_0$ is the Sommer parameter,
from Ref. \cite{ref:precis}. The rho mass, pseudoscalar decay constant,
and pseudoscalar matrix element
 $\lambda_\pi = \langle 0 |  \bar \psi \gamma_5 \psi | PS \rangle$
(all extrapolated to zero quark mass)
 are determined
from simulations on $12^3\times 24$ lattices at the same parameter
values\cite{ref:dh00}, while the infinite volume $\Sigma$ measurement
is from Ref. \cite{ref:TOM_OVER}. The phenomenological estimate
for $\lambda_\pi$, which is equal to $f_\pi m_\pi^2/(m_u+m_d)$, is from
Ref. \cite{ref:s93}. The inferred lattice spacing, shown in the fourth column
of the Table, includes the unknown (but believed to be close to unity)
lattice-to-continuum renormalization factors appropriate to each observable.

The dominant feature of $R_{PS}(x)$ or $R_{PS+S}(x)$ is the pion. Its
contribution is proportional to $\lambda_\pi^2$, or to the fourth power
of the lattice spacing. This is the source of the sensitivity of the
 correlator to $a$.

\begin{table}
\begin{tabular}{|c|l|l|l|}
\hline
quantity & lattice value & continuum & lattice spacing (fm) \\
\hline
$r_0$ & see \protect\cite{ref:precis}    & 0.5 fm            & 0.11 \\
$m_\rho$      & $am_\rho=0.50(2)$ & 770 MeV & 0.13 \\
$f_\pi$       &$af_\pi=0.078(1)$  & 131 MeV & 0.12/$Z_A$ \\
$\lambda_\pi$ & $a^2\lambda_\pi=0.143(3)$ & (470 MeV)${}^2$ & 0.16 $Z_A Z_m$ \\
$\langle \bar\psi\psi\rangle$ & $a^3\Sigma = 0.0039(1)$ & 
(250 MeV)${}^3$ & 0.13 $Z_m^{1/3}$ \\
\hline
\end{tabular}
\caption{Table of observables and inferred lattice spacings.}
\label{tab:latts}
\end{table}

\begin{figure}[thb]
\begin{center}
\epsfxsize=0.8 \hsize
\epsffile{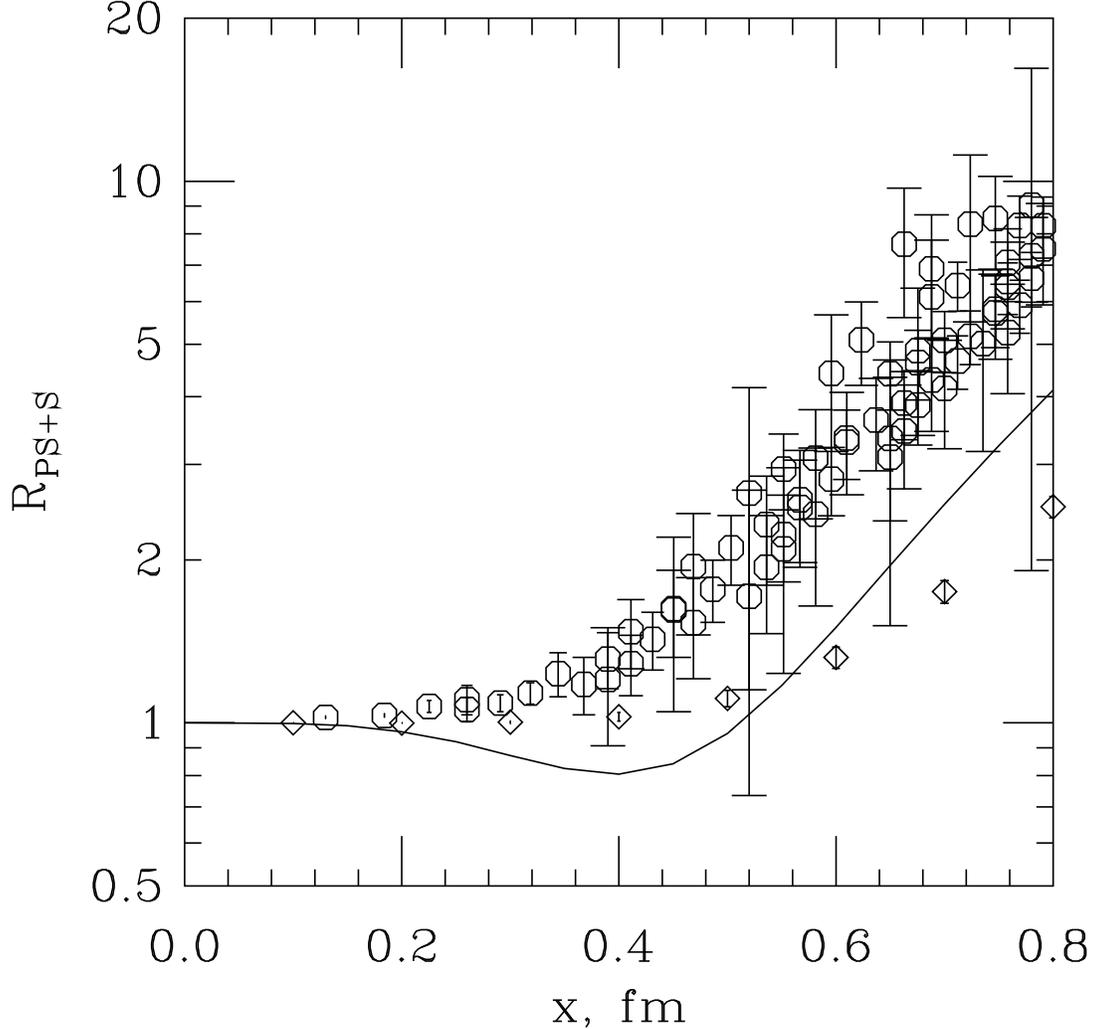}
\end{center}
\caption{
Comparison of the sum of lattice  point-to-point pseudoscalar and
 scalar correlators, extrapolated to zero quark mass (octagons),
predictions of the instanton liquid model (diamonds) and a simple
phenomenological model for $\rho(s)$ (solid line).
}
\label{fig:comppsssha}
\end{figure}

I will somewhat arbitrarily take the lattice spacing to be 0.13 fm in
Fig. \ref{fig:comppsssha}, where I compare the lattice results to
instanton liquid model data from Ref. \cite{ref:ss96}.
The solid line is a very naive phenomenological calculation of
$R_{PS+S}$: I take a spectral function which has a delta-function pion of
weight $\lambda_\pi^2/2$ (with its value from the Table)
 and add in a step function representing
a continuum:
$\rho_{cont}(x) \simeq \theta(s-s_0)$ with $s_0=0.5$ GeV${}^2$.
This parameterization does a very good job of modeling the low quark
mass lattice data, when one inputs the observed lattice values
of pseudoscalar mass and $\lambda_\pi$, as is shown in panel (a)
of Fig. \ref{fig:psscalar}. In Fig. \ref{fig:comppsssha}
the data overshoots the model by about thirty per cent. It will take
simulations at a smaller lattice spacing to disentangle scale violations
from lattice spacing uncertainties, to improve this picture.

\subsection{The vector and axial vector channels}

The potentially interesting physics in these channels involves a comparison
of lattice data to the predictions of the OPE. With my sign conventions,
the coordinate-space quark propagator (with flavor label $a,b$
 in slowly varying external fields is (for references see \cite{ref:SS})
\bee
S(x)^{ab} = {\delta^{ab} \over {2\pi^2}} {{\gamma \cdot x}\over x^4}
+ {\delta^{ab} \over {4\pi^2}} {m \over x^2} + q^a(0) \bar q^b(0) +
{``G'' \over x^2} +\dots
\ee
where the rather schematic ``${G / x^2}$''
represents a  long expression which
 will give the gluon condensate term
in a mesonic correlator. With this quark propagator all the quark mass and
quark condensate dependence in the ``sum''
$R_{V+A}$ correlator cancels out, while the ``difference'' correlator
for massive quarks is
\bee
R_{V-A} = {{m^2 x^2}\over 2} + m \pi^2\langle \bar q q \rangle x^4 + Ax^6
+ \dots
\label{DIFFE}
\ee
and the $A x^6$ term is $A x^6 =
 (\pi^3/9)\alpha_s(x) \langle \bar q q \rangle^2 x^6$
which is estimated by Sch\"afer and Shuryak to be about (x/0.66 fm)${}^6$.

The ``sum'' correlator  sees the gluon condensate
\bee
R_{V+A} = 1 +{{\alpha_s(x)}\over \pi} - {1 \over 384}
\langle g^2 (G_{\mu\nu}^a)^2 \rangle x^4 +
{{2 \pi^3}\over 81}\alpha_s(x) \langle \bar q q \rangle \log(x^2) x^6  +\dots
\label{SUM}
\ee

\begin{figure}[thb]
\begin{center}
\epsfxsize=0.8 \hsize
\epsffile{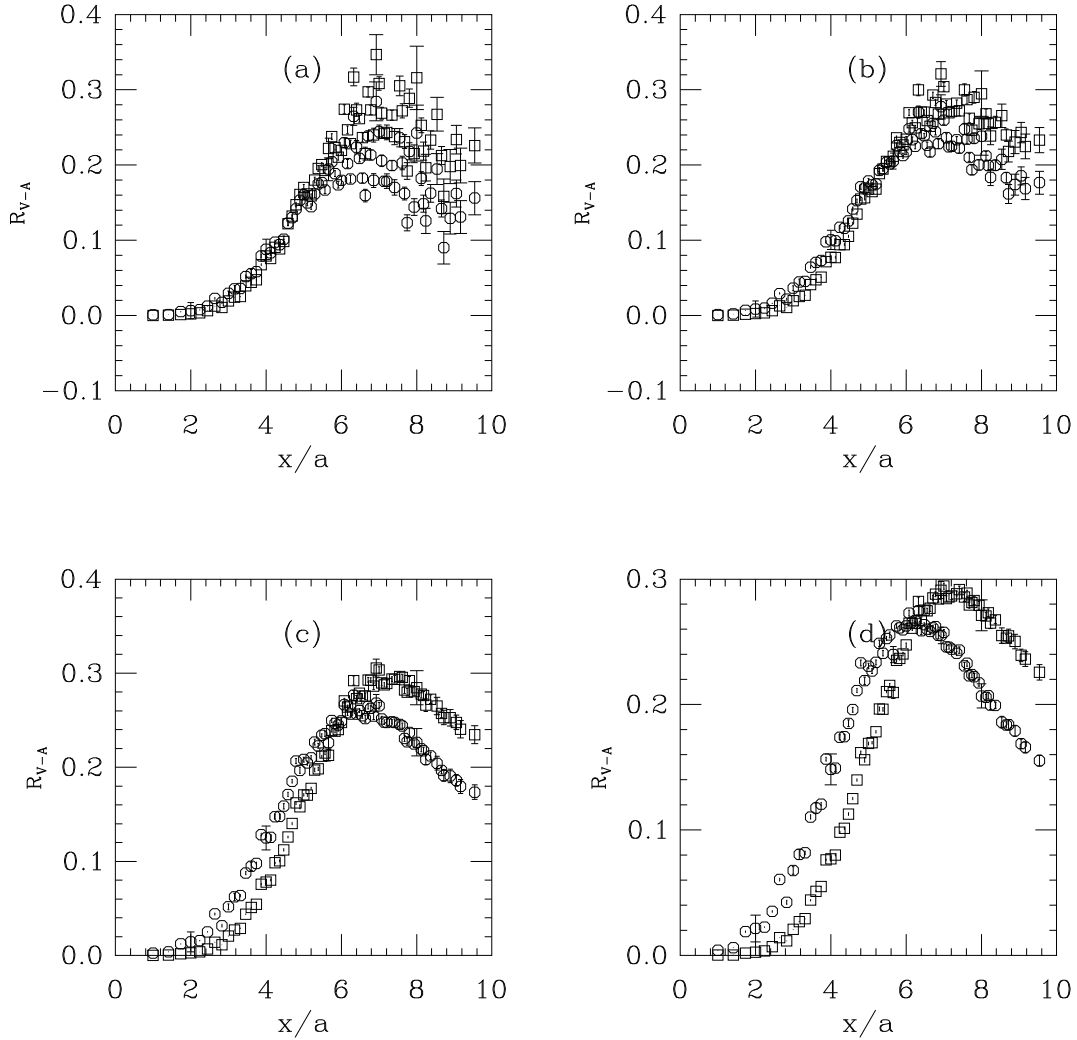}
\end{center}
\caption{
Saturation of the difference of point-to-point vector minus axial
 correlators by  low-lying eigenmodes of $H(0)^2$.
 (a) $am_q=0.01$ ($m_{PS}/m_V\simeq 0.34$);
 (b) $am_q=0.02$ ($m_{PS}/m_V\simeq 0.50$);
 (c) $am_q=0.04$ ($m_{PS}/m_V\simeq 0.61$);
 (d) $am_q=0.06$ ($m_{PS}/m_V\simeq 0.64$).
 Octagons  show the full hadron correlator.
Squares show the contribution from the lowest 10 modes.}
\label{fig:nvpv}
\end{figure}

\begin{figure}[thb]
\begin{center}
\epsfxsize=0.8 \hsize
\epsffile{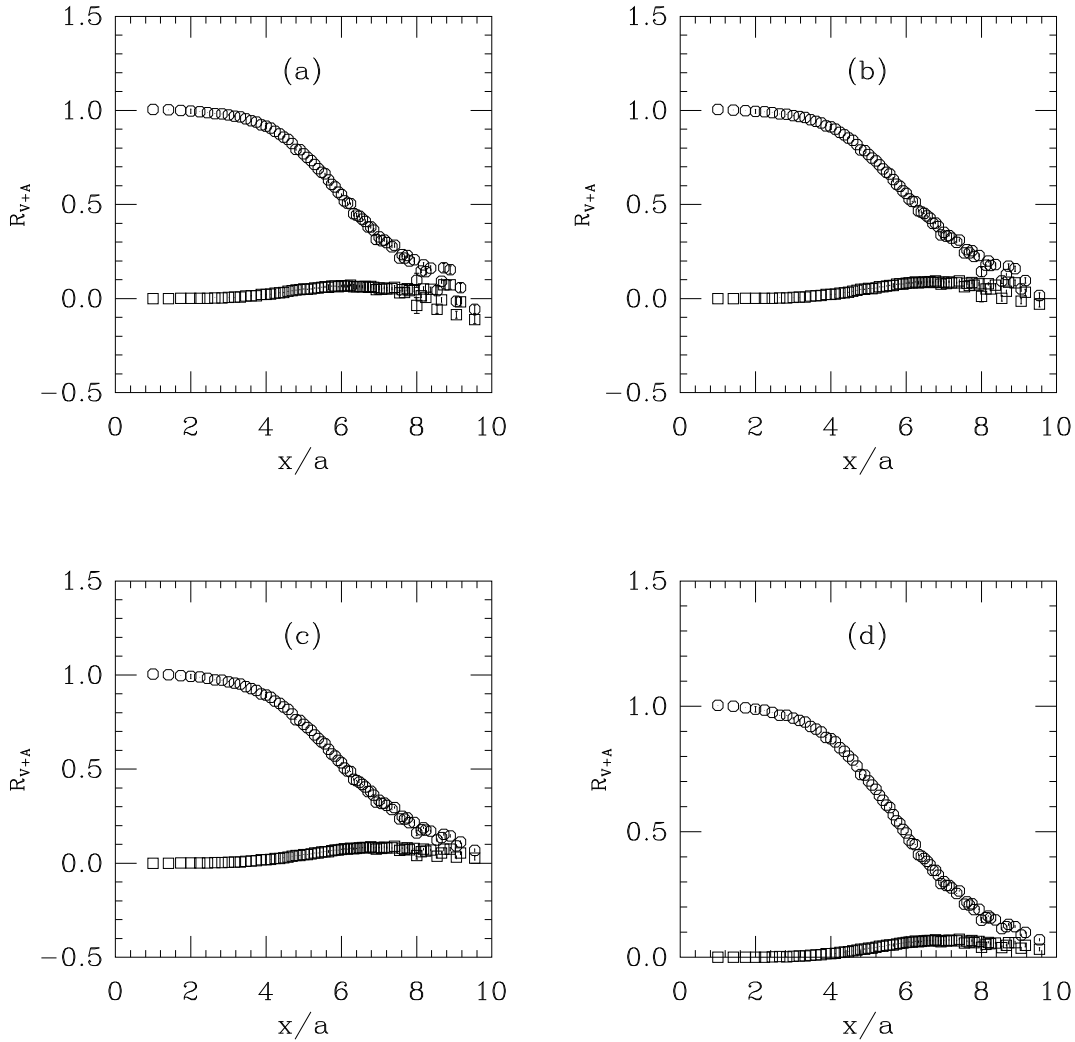}
\end{center}
\caption{
Comparison of the  summed point-to-point vector and axial vector
 correlators from the full quark propagator and with
the propagator built of   low-lying eigenmodes of $H(0)^2$.
 (a) $am_q=0.01$ ($m_{PS}/m_V\simeq 0.34$);
 (b) $am_q=0.02$ ($m_{PS}/m_V\simeq 0.50$);
 (c) $am_q=0.04$ ($m_{PS}/m_V\simeq 0.61$);
 (d) $am_q=0.06$ ($m_{PS}/m_V\simeq 0.64$).
 Octagons  show the full hadron correlator.
Squares show the contribution from the lowest 10 modes.
}
\label{fig:pvpv}
\end{figure}

Figs. \ref{fig:nvpv} and \ref{fig:pvpv} show $R_{V\pm A}$ correlators
from the full simulation and from the lowest ten eigenmodes.
The low modes clearly saturate the ``difference'' correlator
 at $m_{PS}/m_V \le 0.5$.
The ``sum'' correlator is essentially unity, and receives little
contribution from the low eigenmodes.

Now the lattice data is collected at several values of quark mass, all of
which are greater than the physical quark mass, and so to make connection with
phenomenology it is necessary to extrapolate all the data to zero quark mass.
The extrapolation itself can shed light on the OPE prediction: I will assume 
that point-by-point in $x$, $R(x)= r_0 + r_1 m_q + r_2 m_q^2$.
For both correlators, the mass dependence is weak and the extrapolation is
readily performed. However, the coefficients of $m_q$ and $m_q^2$ expected
from the OPE, Eq. \ref{DIFFE}, are only seen at very low $x$, $x/a<4$.

\begin{figure}[thb]
\begin{center}
\epsfxsize=0.8 \hsize
\epsffile{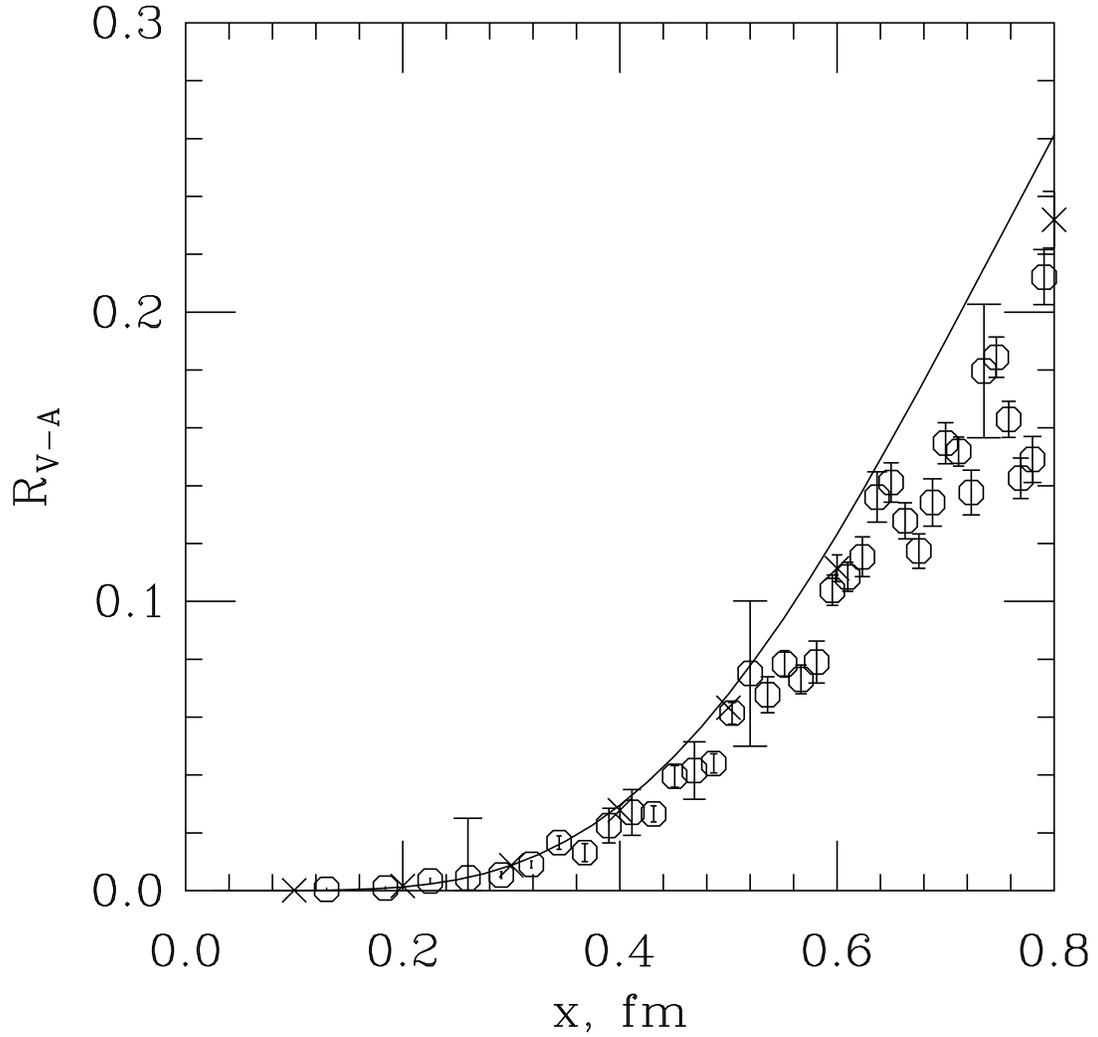}
\end{center}
\caption{
Comparison of the  difference point-to-point vector and axial vector
 correlators from the overlap action (octagons), extrapolated to
zero quark mass, and from
the instanton model of Ref. \protect\cite{ref:ssa1} (crosses)
and ALEPH $\tau-$lepton decay, as extracted by Ref.  \protect\cite{ref:ssa1}
(lines).
}
\label{fig:compnshur}
\end{figure}

\begin{figure}[thb]
\begin{center}
\epsfxsize=0.8 \hsize
\epsffile{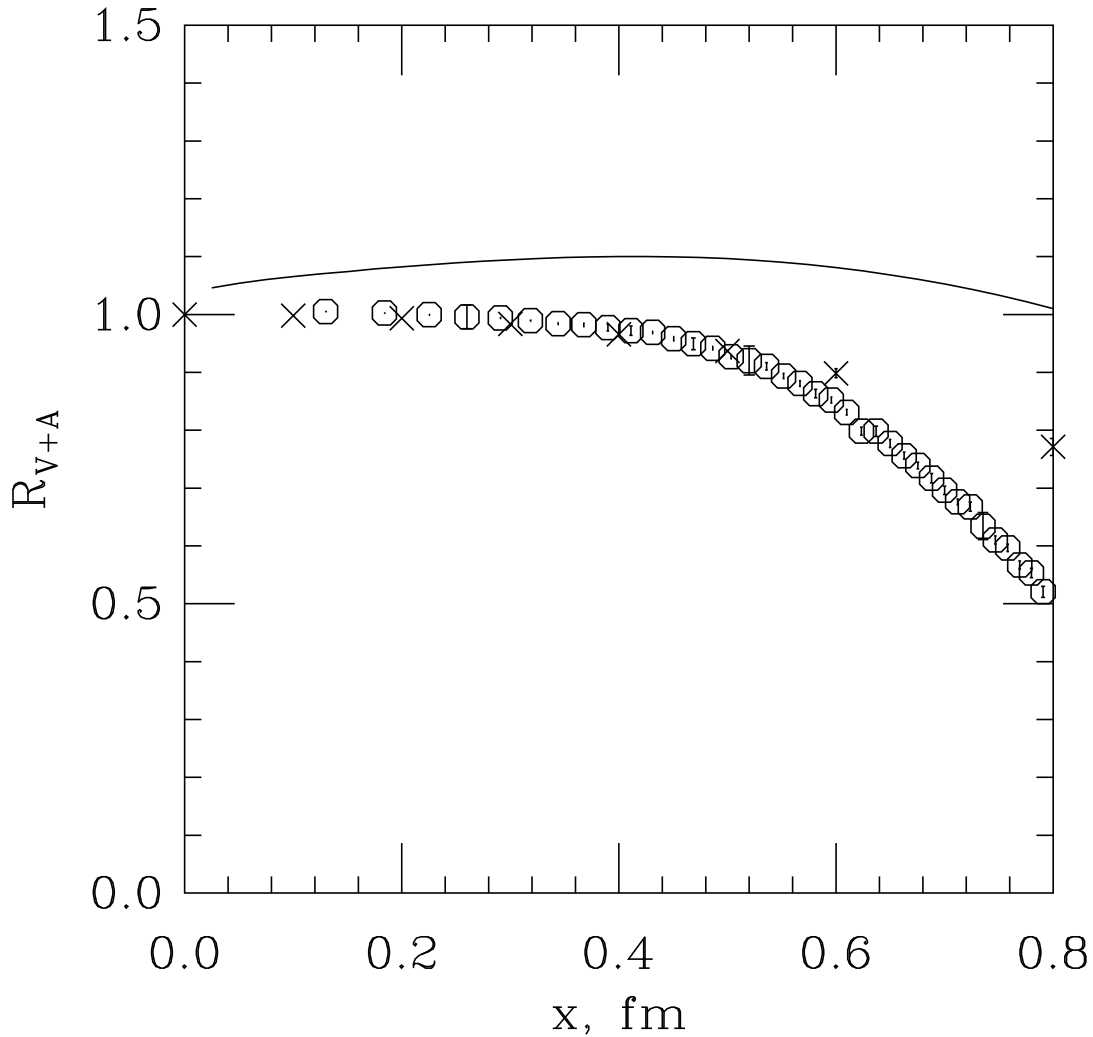}
\end{center}
\caption{
Comparison of the  sum of point-to-point vector and axial vector
 correlators from the overlap action (octagons), extrapolated to
zero quark mass, and from
the instanton model of Ref. \protect\cite{ref:ssa1} (crosses)
and ALEPH $\tau-$lepton decay, as extracted by Ref.  \protect\cite{ref:ssa1}
(lines).
}
\label{fig:comppshur}
\end{figure}

The extrapolated data can be compared to non-lattice results.
Figs. \ref{fig:compnshur} and \ref{fig:comppshur} show
the lattice correlators with
the instanton model of Ref. \cite{ref:ssa1} (crosses)
and ALEPH $\tau-$lepton decay, as extracted by Ref.  \cite{ref:ssa1}
(lines). In these figures I have re-introduced a physical distance
scale (in fm) for the separation $x$ by using the same lattice spacing (0.13 
fm, from the rho mass) as used in the last section. Because the correlator
is such a flat function of $x$, it is not too sensitive to the choice
of lattice spacing.

Lattice Monte Carlo data,
the instanton liquid model, and the extracted tau decay data for $R_{V-A}$
all agree nicely.
This is a channel which has no short distance contributions,
 and the long distance contributions are dominated by instanton-sensitive
eigenmodes.

The lattice data for $R_{V+A}$ also agree with the instanton liquid model out
 to a distance of about 0.6 fm, and then become less attractive.
They undershoot the tau decay data by about five cent.
It is plausible to assume that the latter
 discrepancy is just due to the lack of
 lattice physics below the cutoff scale, and to the fat link used in the
gluon vertex, which smears the action out to about two lattice spacings
(0.26 fm, nominally). It is the same discrepancy as the
 instanton liquid model data, which does not ``naturally'' include the
 perturbative $\alpha_s/\pi$ part of the correlator.
Of course, as I have remarked earlier, I have not computed the lattice to
continuum conversion factor for the vector and axial currents. While it
is unlikely to account for all the discrepancy, it might be the source of
some of it.

The authors of Ref. \cite{ref:ssa1} perform fits to $R_{V\pm A}(x)$
in order to extract the various condensate terms (which will appear
as different powers of $x^n$), as well as to look for $\log x^2$ terms
so characteristic of the operator product expansion.
I tried to do this with the lattice data. Because the lattice data and
instanton liquid model data are so similar, the results will be similar
to those found in Ref. \cite{ref:ssa1}--that is, difficult to
 reconcile with the OPE.

 For example, consider $R_{V-A}$. 
The action itself has order $a^2$ discretization errors, which generate
terms in $R(x)$ proportional to $x^2$. Good fits to the lattice data
 can be found beginning
with any even power of $x$. However, if we take the continuum theory seriously
and consider only fits beginning with $x^6$ term, one can find
 fits for $x < 0.4$ fm or so of the form (x/ 0.6 fm)${}^6$,
quite close to the expectation of the OPE and
the results of Sch\"afer and Shuryak \cite{ref:ss96}.

Fits to $R_{V+A}$ designed to expose the gluon condensate term are
also not very successful. The fit parameters are quite sensitive to
tie range of $x$. Most fits (which range from 0.11 to 0.4 fm or so)
 give a far too large condensate term and the wrong sign for the $x^6$
term.

\section{Conclusions}
Low-lying eigenmodes of the Dirac operator, which we have
previously argued in Ref. \cite{ref:dh00} couple strongly to instantons
and anti-instantons, dominate the nonperturbative part of short
distance correlators for light quark masses in the pseudoscalar,
scalar, vector and axial vector channels.
In the pseudoscalar channel they are responsible for the bulk
of the observed strong attraction.
The zero eigenmodes in the scalar channel contribute a strong repulsion 
at larger distances.
The pseudoscalar plus scalar channel is more attractive on the lattice than in 
the instanton liquid model.
The low modes dominate the vector minus axial vector channel, which 
has no perturbative contribution.
The lattice calculations quantitatively
 reproduce the results of instanton liquid models in the sum and
 difference of vector and
axial vector channels. The difference between the lattice results and
correlation function extracted from tau data in the $V+A$ channel
can plausibly be attributed to short distance physics.
As the quark mass rises, the relative importance of the light modes decreases
in all observed channels.

The lattice simulations could be done much better.
It seems to me that these simulations are rather more sensitive to systematic
effects than they are to statistics. Since one is comparing results to free
field theory, an overlap action with better ``kinetic'' (dispersion relation,
rotational invariance) properties might be a better choice\cite{ref:betterk}.
 Even though these
correlators are called ``short distance,'' a larger volume is needed to 
suppress zero mode effects and to deal with the weak (power law) falloff
of the normalizing free correlators.
(A nonchiral action would be cheaper to simulate, but would suffer from
fatal exceptional configurations at the low values of quark masses needed
to study interesting physics questions.)
Finally, if one wants to look for OPE terms in the correlators it
will be necessary to drop the lattice spacing and convert to links
which smear only a minimal distance--the hypercubic blocking
of Hasenfratz and Knechtli\cite{ref:HK}
 would be a good choice. It seems to me since
most of the physics which is being explored here involves chiral symmetry,
it would be a mistake to look at these correlators with a nonchiral lattice
action.

With the use of a chiral fermion action, lattice calculations may
be poised to say more interesting things about spectral functions in QCD.

\section*{Acknowledgements}
 This work was supported by the
U.~S. Department of Energy. I would like to thank
A. Hasenfratz for numerous conversations, and
 E. Shuryak and T. Sch\"afer for much valuable correspondence and
for providing me with tables of their data.


\end{document}